\documentclass[useAMS,usenatbib]{mn2e}
\usepackage{times}  % This is close to the actual font MNRAS uses

\usepackage{rotating}
\usepackage[T1]{fontenc}

\usepackage{amssymb}
\usepackage{amsmath}
\usepackage{amstext}
\usepackage{float}
\usepackage{graphicx}
\usepackage{caption}
\usepackage{subcaption}
\usepackage{color}

\DeclareRobustCommand{\ion}[2]{%
\relax\ifmmode
\ifx\testbx\f@series
{\mathbf{#1\,\mathsc{#2}}}\else
{\mathrm{#1\,\mathsc{#2}}}\fi
\else\textup{#1\,{\mdseries\textsc{#2}}}%
\fi}

\newcommand{\HI}{\rm H\,{\sc i}}
\newcommand{\HII}{H\,{\sc ii}}

\newcommand{\Ha}{\rm H$\alpha$}
\newcommand{\Hb}{\rm H$\beta$}
\newcommand{\kms}{\rm ~km\,s$^{-1}$}
\newcommand{\nodata}{...}

\newcommand{\MHI}{${\rm M}_{\rm H\,I}$}
\newcommand{\MHII}{${\rm M}_{\rm H\,II}$}
\newcommand{\Mion}{${\rm M}_{\star}$}

\newcommand{\Ms}{${\rm M}_{\rm stars}$}

\newcommand{\Mg}{${\rm M}_{\rm gas}$}

\newcommand{\Mb}{${\rm M}_{\rm bar}$}

\newcommand{\Mdyn}{${\rm M}_{\rm dyn}$}

\newcommand{\Moy}{${\rm M}_{\odot}$\,yr$^{-1}$}

\newcommand{\Mo}{${\rm M}_{\odot}$}

\newcommand{\Te}{$T_{\rm e}$}

\newcommand{\Wabs}{$W_{abs}$}
\newcommand{\abox}{12+log(O/H)}
\newcommand{\lno}{log(N/O)}
\newcommand{\chb}{$c$(H$\beta$)}

\title[Luminous \HII\ complex in GAMA\,J141103.98-003242.3]{The SAMI Galaxy Survey: The discovery of a luminous, low-metallicity \HII\ complex in the dwarf galaxy GAMA\,J141103.98-003242.3}
\author[Richards et al.]
{\parbox{\textwidth}{\raggedright
{S.~N.~Richards$^{1,2,3}$\thanks{E-mail:  \texttt{samuel@physics.usyd.edu.au}},
A.~L.~Schaefer$^{1,2,3}$, 
\'A.~R.~L\'opez-S\'anchez$^{2,4}$, 
S.~M.~Croom$^{1,3}$, 
J.~J.~Bryant$^{2,3}$,
S.~M.~Sweet$^{5}$, 
I.~S.~Konstantopoulos$^{2}$,
J.~T.~Allen$^{1,3}$,
J.~Bland-Hawthorn$^{1}$,
J.~V.~Bloom$^{1,3}$, 
S.~Brough$^{2}$,
L.~M.~R.~Fogarty$^{1,3}$,
M.~Goodwin$^{2}$,
A.~W.~Green$^{2}$,
I.\,-T.~Ho$^{7}$,
L.~J.~Kewley$^{6}$,
B.~S.~Koribalski$^{8}$,
J.~S.~Lawrence$^{2}$,
M.~S.~Owers$^{2}$,
E.~M.~Sadler$^{1,3}$,
R.~Sharp$^{6,3}$}}\vspace{0.4cm} \\
\parbox{\textwidth}{$^{1}$Sydney Institute for Astronomy, School of Physics, University of Sydney, NSW 2006, Australia\\
$^{2}$Australian Astronomical Observatory, PO Box 915, North Ryde, NSW 1670, Australia\\
$^{3}$CAASTRO: ARC Centre of Excellence for All-sky Astrophysics\\
$^{4}$Department of Physics and Astronomy, Macquarie University, NSW 2109, Australia\\
$^{5}$School of Mathematics and Physics, University of Queensland, QLD 4072, Australia\\
$^{6}$Research School of Astronomy and Astrophysics, Australian National University, Cotter Rd., Weston, ACT 2611, Australia\\
$^{7}$Institute for Astronomy, University of Hawaii, 2680 Woodlawn Drive, Honolulu, HI 96822, USA\\
$^{8}$Australia Telescope National Facility, CSIRO Astronomy and Space Science, PO Box 76, Epping NSW 1710, Australia}
}

\begin{document}

\date{Received, 3 July 2014, Accepted 3 September 2014 \vspace{-0.4cm}}

\pagerange{\pageref{firstpage}--\pageref{lastpage}} \pubyear{2014}

\maketitle

\label{firstpage}

%%%%%%%%%%%%%%%%%%%%%%%%%%%%%%%%%%%%%%%%%%%%%%%%
\begin{abstract}
We present the discovery of a luminous unresolved \HII\ complex on the edge of dwarf galaxy GAMA~J141103.98-003242.3 using data from the Sydney-AAO Multi-object Integral field spectrograph (SAMI) Galaxy Survey. This dwarf galaxy is situated at a distance of $\sim100$\,Mpc and contains an unresolved region of \HII\ emission that contributes $\sim70$~per~cent of the galaxy's \Ha\ luminosity, located at the top end of established \HII\ region luminosity functions. For the \HII\ complex, we measure a star-formation rate of $0.147\pm0.041$~\Moy and a metallicity of \abox~$=8.01\pm0.05$ that is lower than the rest of the galaxy by $\sim0.2$~dex. Data from the {\it \HI\ Parkes All-Sky Survey} (HIPASS) indicate the likely presence of neutral hydrogen in the galaxy to potentially fuel ongoing and future star-forming events. We discuss various triggering mechanisms for the intense star-formation activity of this \HII\ complex, where the kinematics of the ionised gas are well described by a rotating disc and do not show any features indicative of interactions. We show that SAMI is an ideal instrument to identify similar systems to GAMA~J141103.98-003242.3, and the SAMI Galaxy Survey is likely to find many more of these systems to aid in the understanding of their formation and evolution. 
\end{abstract}

\begin{keywords}
galaxies: dwarf -- galaxies: starburst -- galaxies: evolution -- (ISM:) H II regions -- techniques: spectroscopic \vspace{-0.8cm}
\end{keywords}

%%%%%%%%%%%%%%%%%%%%%%%%%%%%%%%%%%%%%%%%%%%%%%%%

\section{Introduction}

\begin{table}
\caption{Derived properties for GAMA\,J141103.98-003242.3 and its bright \HII\ complex. Values in the upper section of the table have been derived from previous studies. The values in the lower half of the text have been derived from the SAMI data cube as described in the text. \Ms\, is stellar mass, \Mg\, is neutral gas mass (including helium), \Mb\, is baryonic mass, \Mdyn\, is dynamical mass, \MHII\, is ionised gas mass and \Mion is stellar ionising cluster mass. `Rest of the galaxy' excludes the \HII\ complex.}

\begin{tabular}{@{}lrr@{}}
\hline
  
\multicolumn{3}{c}{GAMA 567676 $=$ GAMA\,J141103.98-003242.3} \\ 
\hline

$\mathrm{RA}$ (J$2000$)          & \multicolumn{2}{c}{14:11:03.98}\\ 
$\mathrm{DEC}$ (J$2000$)         & \multicolumn{2}{c}{-00:32:42.39}\\ 
$z$                              & \multicolumn{2}{c}{$0.0259$}\\ 
$d$ [Mpc]                        & \multicolumn{2}{c}{$106$}\\ 
R$_{e}$                          & \multicolumn{2}{c}{$4\farcs3=1.5$~kpc}\\ 
$i$ [$^{\circ}$]\,$^{\dagger}$   & \multicolumn{2}{c}{$\sim35$}\\

\noalign{\smallskip}

$\log$(\Ms/\Mo)                  & \multicolumn{2}{c}{$8.52\pm0.13$}\\       
$\log$(\Mg/\Mo)                  & \multicolumn{2}{c}{$\sim9.62$}\\ 
$\log$(\Mb/\Mo)                  & \multicolumn{2}{c}{$\sim9.66$}\\
        \Mg/\Ms                  & \multicolumn{2}{c}{$\sim11.2$}\\
$v_{{\rm rot},{\rm \sc H\,I}}$ [\kms]            & \multicolumn{2}{c}{$\sim55$}\\   
$\log$(\Mdyn/\Mo)                & \multicolumn{2}{c}{$\sim9.74$}\\
 
\noalign{\smallskip}

\Ha-SFR [\Moy]  (GAMA)           &\multicolumn{2}{c}{$0.121$}\\ 
$FUV$-SFR [\Moy]                 &\multicolumn{2}{c}{$0.58\pm0.24$}\\

\noalign{\smallskip}
\hline
\noalign{\smallskip}
       &    \HII\ complex  &    Rest of the galaxy \\ 
\hline
\noalign{\smallskip}

EW(H$\alpha$) [\AA]  &  $451\pm11$ & $34.8\pm3.2$ \\
EW(H$\beta$) [\AA]  &  $137\pm4$ & $6.8\pm1.6$ \\
 
$F$(H$\alpha$) [$10^{-16}$\,erg\,s$^{-1}$\,cm$^{-2}$]\,$^{\ddagger}$   &$138\pm38$  &    $54\pm15$  \\
$L$(H$\alpha$) [$10^{39}$\,erg\,s$^{-1}$] & $18.5\pm5.2$ & $7.3\pm2.0$ \\
 
$\log$(\MHII/\Mo)       &  $5.58\pm0.11$    &  $5.02\pm0.11$   \\

$\log$(\Mion/\Mo)     &  $5.81$ & $6.53$   \\

\noalign{\smallskip}

Equivalent O7V stars    &  $1360$ &   $538$ \\
Age of most recent starburst [Myr]  &  $4.8\pm0.2$ & $7.9\pm0.4$ \\

\noalign{\smallskip}

$E(B-V)$  [mag]        & $0.24\pm0.01$   &  $0.21\pm0.04$ \\
\abox                  & $8.01\pm0.05$   &  $8.18\pm0.11$ \\
\lno                   & $-1.43\pm0.06$  &  $-1.51\pm0.10$ \\

\noalign{\smallskip}

\Ha-SFR [\Moy]      &$0.147\pm0.041$ & $0.058\pm0.016$\\

\hline
\end{tabular}\\
\scriptsize{$^{\dagger}$ Derived from the ellipticity on GAMA's Sersic fit to the $r$-band photometry.\\
$^{\ddagger}$ Error is currently dominated by a $28$~per~cent error in SAMI's absolute flux calibration, and in all values derived from it (see \citet{All14} for more details).}
\label{tab:Summ}
\end{table}

%%%%%%%%%%%%%%%%%%%%%%%%%%%%%%%%%%%%%%%%%%%%%%%%

Galaxies are thought to have formed hierarchically from the agglomeration of smaller structures. While dwarf galaxies (DGs) occupy the lower end of the galaxy mass function, they are a critical population for the understanding of galaxy evolution as they represent the fundamental units of galaxy formation in the early universe. Over the years there have been varying definitions as to what is classified as a DG \citep{Hod71,Tam94,Mat98}. For the purpose of this work, we define a DG as a galaxy with a stellar mass < $10^9$\,\Mo. Despite DGs being the most numerous galaxies in the Universe \citep{Fon06}, their low stellar masses frequently result in a low surface brightness. Due to the \emph{Malmquist bias}, extensive studies of DGs have therefore only been carried out in the Local Group \citep[e.g.][]{Hun93,Mat98,Beg08,Tol09} and within the Local Volume \citep[e.g.][]{Ken08,Kirby+08,Bouchard+09,Dal09,Young+14}. Few analyses of DGs extend beyond $\sim10$~Mpc distance. These studies have been typically focussed on Blue Compact Dwarf galaxies \citep[BCDGs;][]{Gil03, Izo04,LSE08,Cai09, Hun10, Karthick+14} that show intense star-forming knots within a clumpy \Ha\ morphology. Analyses of \HII\ regions in dwarf galaxies are important for understanding their star-formation history. Observationally, emission-line diagnostics \citep[e.g.][]{KD02,Dop06,Dop13} or theoretical evolution synthesis models \citep[e.g.][]{L99,BC03,Molla+09} can be used to constrain the physical parameters of extragalactic \HII\ regions and their host galaxies.

The \Ha\ luminosity function of \HII\ regions in BCDGs covers the range $10^{36}$--$10^{41}$\,erg\,s$^{-1}$, and seems to follow those of larger nearby galaxies \citep{Oey98,You99,Bra06}. \HII\ regions with \Ha\ luminosities on the order of $>10^{40}$\,erg\,s$^{-1}$ (the top percentile of the \HII\ region luminosity function) are of particular interest as they are the hosts of the most extreme star formation events, which drive the evolution of galaxies. For comparison, the famous \HII\ complex 30~Doradus (NGC\,2070) in the Large Magellanic Cloud (LMC) has an \Ha\ luminosity of $1.5~\times~10^{40}$\,erg\,s$^{-1}$ \citep{Ken84}, a metallicity of \abox$~=8.33\pm0.02$ \citep{Peimbert03}, and it is currently forming a massive star cluster \citep{Bosch09}. {\rm SBS 0335-052E, II Zw 40 and J1253-0312} \citep{Pus04,MK06,Gus11} are BCDGs that occupy the very extreme end of the \HII\ region luminosity function with a \Ha\ luminosities $>10^{41}$\,erg\,s$^{-1}$. In a massive star cluster the first supernovae will occur roughly a million years after the initial burst of star formation and may expel a large fraction of the gas within. Therefore a sustained star formation rate (SFR) of order 0.1\,\Moy\ is required for the formation of a cluster of $\sim10^{5}$\,\Mo.

Some works have shown that galaxy interactions trigger strong star-formation activity in dwarf galaxies \citep[e.g.][]{Kor09}. The multiwavelength analysis of BCDGs performed by \citet{ALS10} found that the majority of them were clearly interacting or merging with low-luminosity dwarf objects or \HI\ clouds. The interacting features were only detected by deep optical spectroscopy and detailed multiwavelength analysis, which includes a study of the kinematics and distribution of the neutral gas. Indeed, many times the disturbances were found when examining the neutral gas, as was the case in the dwarf galaxies NGC\,1705 \citep{Meu98}, NGC\,625 \citep{Can04}, NGC\,1569 \citep{Muh05}, IC\,4662 \citep{vEy10} and  NGC\,5253 \citep{LSK+12}. This is often the case because low mass companions often have a high gas fraction and are therefore visible in \HI\ emission. Moreover, aperture synthesis observations of spectral lines are usually performed at high spectral resolution, allowing kinematic maps of these sources to be produced. An example of a merger-induced star formation event in a dwarf system is SBS\,1319+579 \citep{LSE09}, where long slit spectroscopy revealed differing velocity components for the merging systems. 

Luminous \HII\ regions can also be created stochastically by processes internal to a galaxy. For example, star formation could either be triggered by density waves propagating through an irregular distribution of \HI\ \citep{Ger80} or solely by the gravitational collapse of a gas cloud within the galaxy disc \citep{Lad08}. \citet{Bau13} analysed the SFRs and specific-SFRs (SSFRs) of low-mass ($<10^{10}$\,\Mo\,) galaxies within the GAMA-I survey and concluded that their star formation histories (SFHs) require stochastic bursts of star formation superimposed onto an underlying exponentially declining SFH. This is enhanced in low-mass galaxies as they have fewer individual star forming regions compared to more massive galaxies \citep{Lee09}.

Clumpy star-forming galaxies at higher redshift ($z\sim1$--$2$), so called `clump-clusters' \citep{Elm04}, host star forming clumps (regions) with diameters on the scale of a few kpc \citep{Wis12}. Local dwarf-irregular galaxies have been used as analogues for these higher redshift galaxies, as they can be seen as currently undergoing the same evolutionary phase. The difference is the timescale over which these large star forming regions and the host galaxy evolve; inversely proportional to their stellar mass. Today most of these clump-clusters have evolved into smooth discs, leaving dwarf irregulars as a visible example of this phase \citep{Elm09}. 

Observationally, extragalactic \HII\ regions are best identified with spatially-resolved spectroscopy or emission-line imaging. Broad-band photometry using {\rm g-r vs. r-i} colour-colour diagrams \citep{Car09,Izo11} or narrow-band \Ha\ imaging can be used to identify \HII\ regions with strong emission lines, but they are unable to obtain line ratios and velocity information. Integral field spectroscopy (IFS), on the other hand, can trace dynamics, metallicities and gas processes through line velocities and abundances, which can help to identify the triggers for star formation. Besides some analyses of individual objects \citep[e.g.][]{James+09,James+10,Monreal-Ibero+10,LS+11,Perez-Montero+11}, few IFS surveys, with the exception of the {\it Small Isolated Gas Rich Irregular Dwarf galaxy survey} \citep[SIGRID, 83 DGs;][]{Nic11}, the  {\it IFS - Blue Compact Galaxy survey}  \citep[IFS-BCG, 40 DGs;][]{Cai12}, and the {\it Choirs survey} \citep[Choirs, 38 DGs;][]{Swe13}, have been targeting dwarf galaxies. As all of these works used a single Integral Field Unit (IFU) instrument, the time taken to gather their data has been lengthy. The quoted number of targets for each survey is the proposed quantity, with no survey complete at the time of writing.

The SAMI Galaxy Survey \citep{Cro12, Bry14b, Sha14, All14}, which uses the Sydney-AAO Multi-object Integral field spectrograph (SAMI), deployed at prime focus on the $3.9$\,m Anglo-Australian Telescope (AAT), started observations in early 2013, and seeks to obtain spatially resolved spectra for $\sim3400$ galaxies over 3 years. This survey has the ability to identify and analyse \HII\ complexes in galaxies with a large range of stellar masses ($7.4\lesssim\log$(M/\Mo) $\lesssim12.7$). The survey is volume-limited with steps in redshift (see \citet{Bry14b} for details of the target selection). Consequently $\sim400$ galaxies in the sample will have stellar masses $<10^9$\,\Mo. Upon visually checking $70$ DGs that have already been observed with SAMI, GAMA\,J141103.98-003242.3 stood out due to an unusually luminous \HII\ complex located at the edge of the galaxy, which was strikingly visible in its \Ha\ map, but barely seen in the continuum maps. This galaxy and its luminous \HII\ complex are the focus of the work presented here.

In Section $2$ we present existing data on GAMA\,J141103.98-003242.3; Section $3$ we describe the SAMI observations that are the focus of this paper; Section $4$ we detail measurements made from the SAMI data; Section $5$ we discuss likely mechanisms that are driving the evolution of GAMA\,J141103.98-003242.3. In this paper we assume the standard $\Lambda$CDM cosmology with $\Omega_{m}=0.3$, $\Omega_{\Lambda}=~0.7$ and $H_0$ = $73$\,km\,s$^{-1}$\,Mpc$^{-1}$.

%%%%%%%%%%%%%%%%%%%%%%%%%%%%%%%%%%%%%%%%%%%%%%%%

\begin{figure}
\centering
\includegraphics[width=\linewidth]{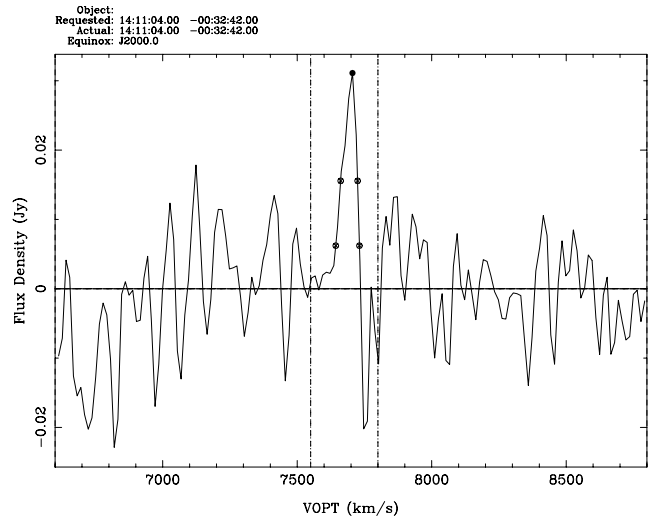}
\caption{The HIPASS radio spectrum in the direction of GAMAJ141103.98-003242.3. This spectrum shows evidence for a $21$\,cm neutral hydrogen emission feature at the redshift of the galaxy with a flux of $1.4$\,Jy\,km\,s$^{-1}$ (central dot), a $50$~per~cent velocity width of $63$\,km\,s$^{-1}$ (inner two dots), and an $80$~per~cent velocity width of $88$\,km\,s$^{-1}$ (outer two dots). The FWHM of the Parkes gridded beam is $15.5$~arcmin. The typical $rms$ noise in HIPASS is $13$~mJy\,beam$^{-1}$. The measured \HI\ peak flux is $\sim2.5\,\sigma$, the integrated flux is $\sim4\,\sigma$. The HIPASS channel width is $13.2$~km\,s$^{-1}$; the velocity resolution is 18~km\,s$^{-1}$. VOPT is the velocity scale of the HIPASS data with respect to the solar barycentre in the usual optical (cz) convention. }
\label{fig:HIPASS}
\end{figure}

%%%%%%%%%%%%%%%%%%%%%%%%%%%%%%%%%%%%%%%%%%%%%%%%

\begin{figure*}
\centering
\includegraphics[width=16.8cm]{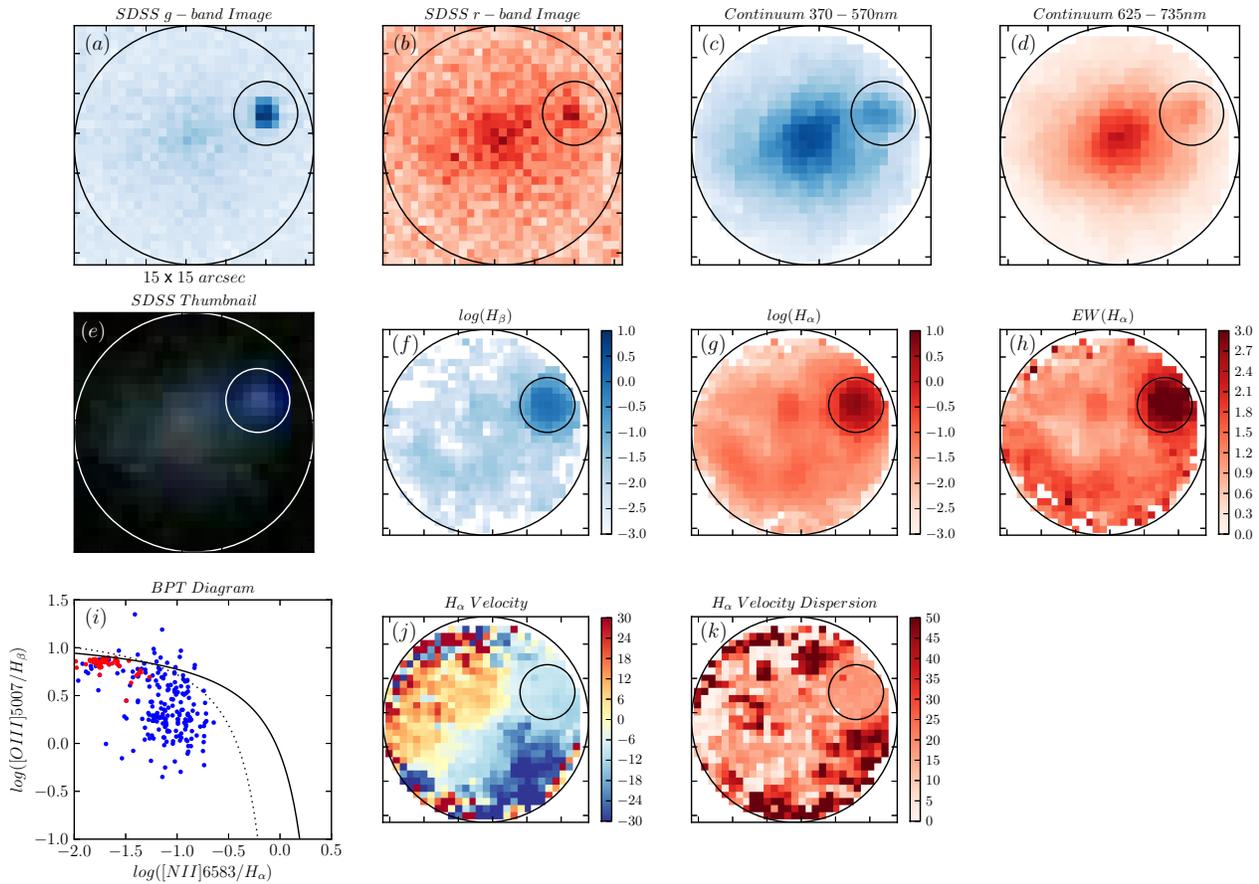}
\caption{Diagnostic plots for GAMAJ141103.98-003242.3. (a) and (b) are the SDSS g- and r-band images. (c) and (d) are the LZIFU fitted continuum maps. (e) is the SDSS 3-colour thumbnail. (f) and (g) are the \Hb\ and \Ha\ maps respectively (normalised to $10^{-16}$\,ergs\,s$^{-1}$\,cm$^{-2}$). (h) is the log(EW(\,\Ha\,)) map in units of \AA. (i) is the BPT diagram where the spaxels are colour-coded such that the red points are the spaxels within the \HII\ complex, and the blue spaxels are the rest of the galaxy. The black solid and dotted lines represent the star formation limits as prescribed by \citet{Kew01} and \citet{Kau03} respectively. (j) is the \Ha\ velocity field in km\,s$^{-1}$. (k) is the \Ha\ velocity dispersion in km\,s$^{-1}$. In all images, the box is $15$~$\times$$15$~arcsec ($0.5$-arcsec per pixel), the large black circle enclosing the galaxy has a diameter of $15$~arcsec, and the small black circle (centred about the \HII\ complex) has a diameter of $4$~arcsec. In all images, North is up and East is left.\vspace{-0.4cm}}
\label{fig:Summ}
\end{figure*}

%%%%%%%%%%%%%%%%%%%%%%%%%%%%%%%%%%%%%%%%%%%%%%%%
\vspace{-0.4cm}
\section{Existing data on GAMA\,J141103.98-003242.3}

In this section we introduce the properties of GAMA\,J141103.98-003242.3 as determined by prior observations. It was observed by GAMA \citep{Dri11,Hopkins+13}, who find a spectroscopic redshift of $z=0.0259$ ($106$\,Mpc), and from the SDSS photometry find a Sersic radius of $4.26$~arcsec \citep[$1.98$\,kpc;][]{Kel12} and a stellar mass of $\log$(\Ms/\Mo)~$=8.52\pm0.13$ \citep{Tay11}.  

The main properties of this galaxy are tabulated in Table~\ref{tab:Summ}. GAMA also calculates an aperture-corrected star-formation rate (SFR) of $0.121$\,\Moy\ \citep{Gun13}. SDSS optical \citep[Petrosian $r$-band magnitude of $18.16$;][]{Aba09} and GALEX UV photometry \citep{Mor07} are also available for this galaxy. The measured $FUV$ magnitude and $FUV$--$NUV$ colour are $20.63\pm0.05$ and $0.29\pm0.05$ respectively, and applying the \citet{Sal07} calibration to their respective fluxes, a $FUV$-based SFR of $0.58\pm0.24$\,\Moy\ is found.

This galaxy has a  $4\sigma$ detection of the \HI\ 21-cm line using the {\it \HI\ Parkes All-Sky Survey} \citep[HIPASS;][]{Bar01} measured over multiple channels, as shown in Fig.~\ref{fig:HIPASS}. This translates into an \HI\ mass of $\log$(\MHI/\Mo)$~\sim9.50$ and a $50$~per~cent velocity width of 63\,\kms.  Neglecting the molecular gas (which is expected to be very low in low-metallicity galaxies), and accounting for helium ($25$~per~cent), the total neutral gas mass is $\log$(\Mg/\Mo)$~\sim9.62$. Hence, the baryonic mass is $\log$(\Mb/\Mo)$~\sim9.66$ and the gas-to-stellar mass ratio \Mg/\Ms$~\sim11.2$. Assuming that the motion of the neutral gas is due to disc rotation and considering an apparent inclination angle of $i\sim35^{\circ}$ (as measured from the ellipticity values from the GAMA Sersic fit), we estimate a maximum rotational velocity amplitude of $v_{{\rm rot},{\rm \sc H\,I}}\approx55$\,\kms. Considering that the neutral gas may be extended at least up to 4 times the optical size of the galaxy \citep[$4\times R_e = 7.9$\,kpc; e.g.][]{War04}, this yields to a total dynamical mass of $\log$(\Mdyn/\Mo)$~\sim9.74$. Therefore, this dwarf galaxy possesses at least a mass of \mbox{$\log$(M/\Mo)$~\sim8.97$} in the form of dark matter. It should be noted that the HIPASS beam FWHM is 15~arcmin. However, given that the redshift of the \HI\ detection closely matches the optical redshift of the galaxy ($v_{opt}=7765$~km\,s$^{-1}$, $v_{\rm \sc H\,I}\approx7675$~km\,s$^{-1}$) and that there are no catalogued galaxies at the same redshift, it is highly likely that the detected neutral gas is associated with GAMA\,J141103.98-003242.3.

%%%%%%%%%%%%%%%%%%%%%%%%%%%%%%%%%%%%%%%%%%%%%%%%

\begin{figure*}
\centering
\begin{subfigure}[b]{0.95\textwidth}\includegraphics[width=\textwidth]{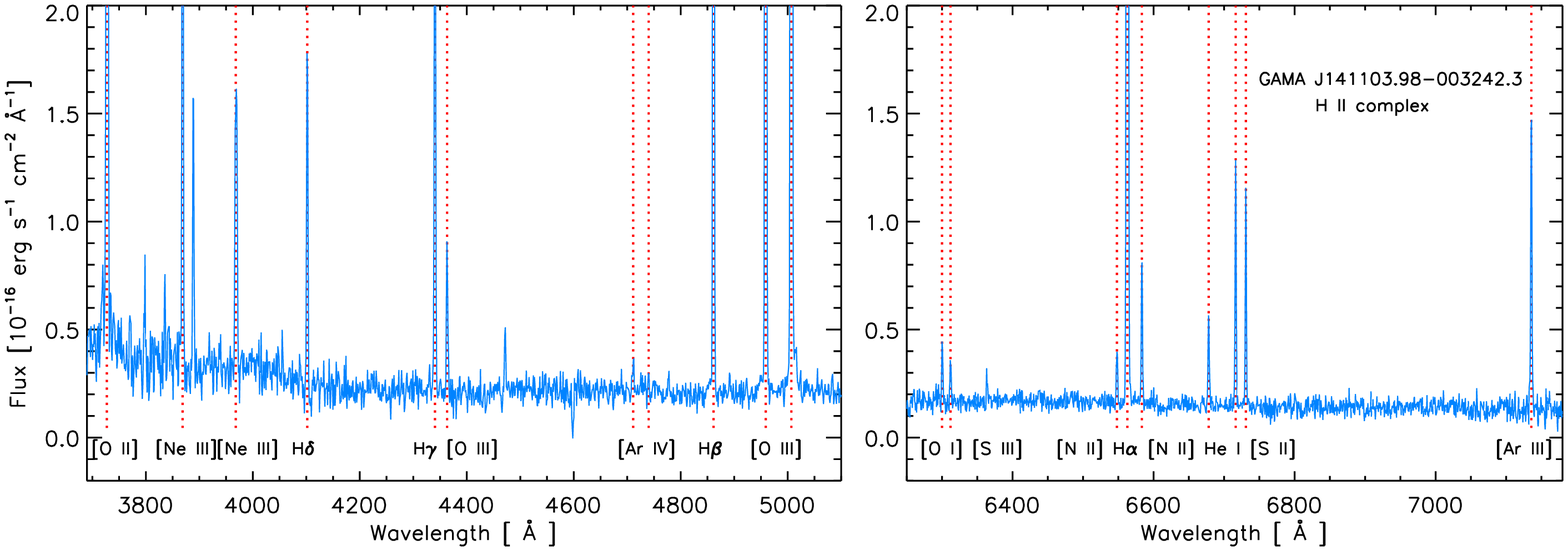}
\end{subfigure}
\begin{subfigure}[b]{0.95\textwidth}\includegraphics[width=\textwidth]{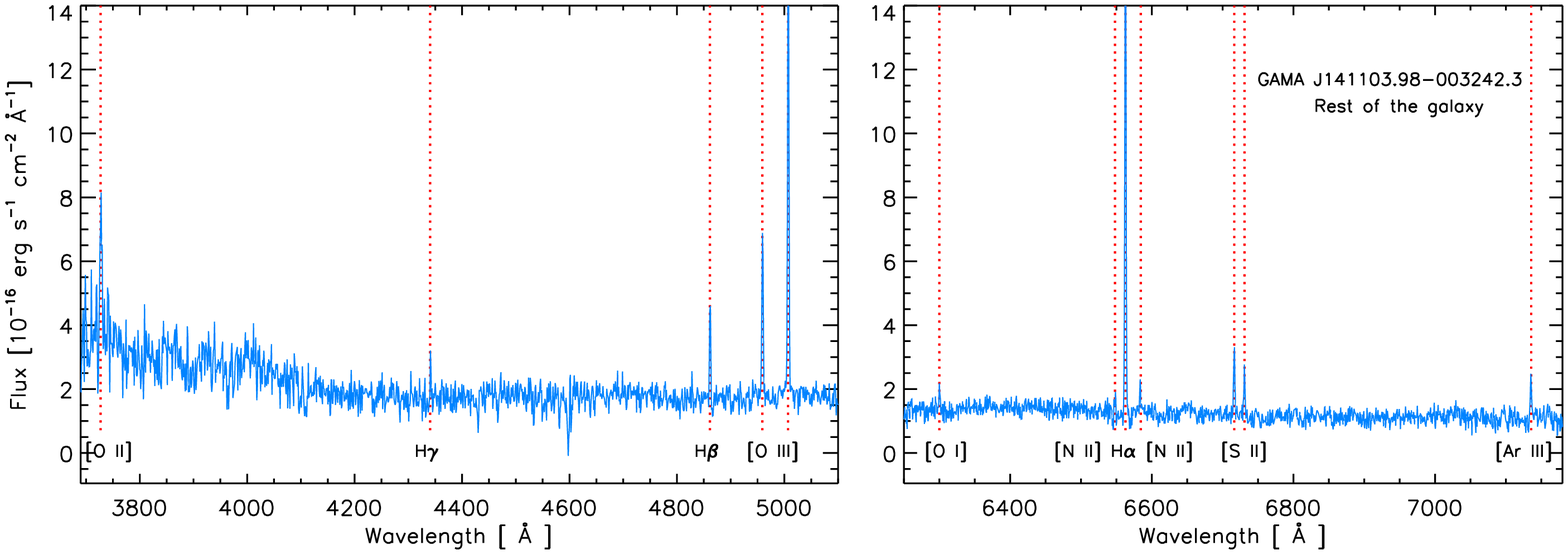}
\end{subfigure}
\caption{Binned optical spectrum of the \HII\ complex (top) and the rest of the galaxy (bottom) using SAMI data. `Rest of the galaxy' excludes the \HII\ complex. The main emission lines are identified by red dotted lines. The wavelength scale is expressed in the rest frame of the galaxy. The red spectrum only shows the $\lambda\lambda\,6250$--$7175$ range. Note that the flux scales in both the blue and red spectra have been truncated in order to accentuate the weaker emission lines. }
\label{fig:spectrum}
\end{figure*}

%%%%%%%%%%%%%%%%%%%%%%%%%%%%%%%%%%%%%%%%%%%%%%%%
\vspace{-0.4cm}
\section{Observations and data reduction}

The data on which this analysis is based were obtained with the SAMI instrument. SAMI deploys $13$ hexabundles \citep{Bla11, Bry14a} over a $1$ degree field at the Prime Focus of the AAT. Each hexabundle is made up of $61$ optical fibres, circularly packed, with each having a core size of $1.6$~arcsec, resulting in a hexabundle field of view (f.o.v) of $15$~arcsec diameter. All $819$ fibres ($793$ object fibres and $26$ sky fibres) feed into the AAOmega spectrograph \citep{Sha06}, configured to a wavelength coverage of $370$--$570$\,nm with $R=1730$ in the blue arm, and $625$--$735$\,nm with $R=4500$ in the red arm. Twelve galaxies are observed in one pointing, along with a secondary standard star that is used for telluric correction and PSF sampling. All frames have a flux zero-point respective to a stand-alone primary standard star observation in one of the bundles, meaning variable sky conditions between the primary standard star observation and the galaxy frames is not taken into account. Due to this, in addition to SAMI's data reduction pipeline, which is still in development, preliminary absolute flux calibration is found to be consistent across the survey data set to $\sim6$~per~cent. However, a residual baseline offset uncertainty between frames taken in variable conditions is not yet accounted for and hence we estimate a $28$~per~cent uncertainty in this early survey data \citet[see][for more details]{All14}.

GAMA\,J141103.98-003242.3 was observed using SAMI on 2013 March 15, with some high level clouds, and a seeing of $\sim2$~arcsec. A seven point dither was performed to achieve near-uniform spatial coverage, with $1800$\,s exposure time for each frame, totalling $3.5$\,h. 

The raw data were reduced using the AAOmega data reduction pipeline, {\small 2dfDR}{\sc}\footnote{{\small 2dfDR}{\sc} is a public data reduction package managed by the Australian Astronomical Observatory, see http://www.aao.gov.au/science/software/2dfdr.}, followed by full alignment and calibration through the SAMI Data Reduction pipeline (see \citet{Sha14} for a detailed explanation of this package). The resulting product of SAMI-DR is a data cube for each of the blue and red observations, both of which have been flux calibrated and corrected for differential atmospheric refraction. 

To obtain the line, velocity and velocity dispersion maps of each galaxy (subtracted for instrumental broadening), the reduced cubes were fitted using a new IFU line fitting package called {\small LZIFU}{\sc} (see Ho et al. \emph{in prep} for a detailed explanation of this package). This software makes use of the {\small pPXF}{\sc} \citep{Cap04} stellar template fitting routine as well as the {\small MPFIT}{\sc} library \citep{Markwardt09} for estimating emission line properties. {\small LZIFU}{\sc} has the ability to perform multi-component Gaussian fits to each emission line, though in the case of GAMA\,J141103.98-003242.3 only a single component was fitted as a multicomponent fit did not give a significant improvement in the reduced chi-squared. It was from the product of {\small LZIFU}{\sc} that this system was identified. To perform detailed analysis, {\small IRAF}{\sc}\footnote{{\small IRAF}{\sc} ({\it Image Reduction and Analysis  Facility}) is distributed by NOAO which is operated by AURA Inc., under cooperative agreement with NSF.} software was used to analyse the summed, aperture-extracted 1D spectra. Line fluxes and equivalent widths were measured by integrating all the flux in the line between two given limits and over a local continuum estimated by visual inspection of the spectra. Visual inspection of the spectra is needed to get a proper estimation of the adjacent continuum and hence a reliable line flux estimation when emission lines are faint. The errors associated with the line flux measurements were estimated by remeasuring the noise ($rms$) in the adjacent continuum of each emission line. 

%%%%%%%%%%%%%%%%%%%%%%%%%%%%%%%%%%%%%%%%%%%%%%%%
\vspace{-0.4cm}
\section{Results}

The products of LZIFU give spatially-resolved information regarding the distribution of ionised gas in GAMA\,J141103.98-003242.3. From the data collected, maps were generated of the H$\alpha$, H$\beta$, [\ion{N}{ii}]~$\lambda\,6583$, [\ion{O}{iii}]~$\lambda\,5007$ and [\ion{O}{ii}]~$\lambda\,3727$  distribution, as well as the line-of-sight velocity field; these are displayed in Fig.~\ref{fig:Summ}. A bright emission region, located $\sim5.1$~arcsec ($1.2\,R_{e}=~2.6$\,kpc) north-west of the galaxy centre, clearly appears in the \Ha\ map -- at the same velocity as the galaxy. This is interpreted as an off-centre, luminous star forming complex within the dwarf galaxy. A 2-dimensional Gaussian fit to the \HII\ complex in the SDSS $r$-band image gives a FWHM of $1.2$~arcsec. This places an upper limit of $\sim600$\,pc on its diameter, implying that the \HII\ complex is unresolved by both SDSS and the SAMI Galaxy Survey. 

From the emission line fits, we are able to produce maps of the velocity and velocity dispersion in the interstellar medium of GAMA\,J141103.98-003242.3. These maps are displayed in Fig.~\ref{fig:Summ}. The gas velocity map is typical of a disc rotating about a northwest-southeast axis with a maximum amplitude of approximately $25$\,km\,s$^{-1}$, which is approximately half of the $v_{{\rm rot},{\rm \sc H\,I}}$ estimated from the 21-cm \HI\ line. The velocity map displays a smooth gradient across the entire field of view, suggesting that the \HII\ complex is co-rotating with the galaxy. 

Areas of the data cube corresponding to the \HII\ complex and the remainder of the galaxy were binned to obtain measurements with reduced random errors. A circle of radius $2$~arcsec, centred on the peak of the \Ha\ intensity, was used to delimit the zone of the cube regarded as part of the \HII\ complex. The binned spectrum of the \HII\ complex and the rest of the galaxy are shown in Fig.~\ref{fig:spectrum}, with all emission lines measurements performed manually (as discussed in Section 3).

In the binned spectrum of the \HII\ complex we also detect emission lines of  \ion{He}{i}, \ion{He}{ii}, [\ion{O}{i}], [\ion{S}{ii}], [\ion{S}{iii}], [\ion{Ne}{iii}], [\ion{Ar}{iii}] and [\ion{Ar}{iv}], as well as the auroral [\ion{O}{iii}]~$\lambda\,4363$ line and many \ion{H}{i} Balmer lines. The complete list of emission lines detected is shown in Table~\ref{tab:lines}, which also includes the dereddened line intensity ratios with respect to $I$(\,\Hb\,)$~=100$. The presence of high-excitation ions such as [\ion{Ar}{iv}] or \ion{He}{ii} indicates the high degree of ionization of the gas. We have followed the same prescriptions described in \citet{LSE09} to determine the physical conditions (electron density, electron temperature, ionization degree, and reddening) and chemical abundances of this \HII\ complex. The reddening coefficient, \chb, was computed using 3 pairs of \ion{H}{i} Balmer lines assuming their theoretical ratios for case B recombination given by \citet{SH95} and a  \citet{Cardelli89} extinction law. We assume the same \Wabs\ for all Balmer lines and contributions other than extinction and stellar absorption are negligible, with both \Wabs\ and \chb\ being simultaneously optimised to best match the observed Balmer pair ratios. A full description of this method is given in \cite{MB93} and \cite{LSE09}. This method, applied to each detectable Balmer line, provides a consistent value of \chb\,$=0.34\pm0.02$. The electron density, computed via the [\ion{S}{ii}] $\lambda\lambda\,6716,6731$ doublet, is $n_e=140\pm30$\,cm$^{-3}$. 

The detection of the [\ion{O}{iii}]~$\lambda\,4363$ line allows us to compute the electron temperature via the  [\ion{O}{iii}] ($\lambda\,4959$+$\lambda\,5507$)/$\lambda\,4363$ ratio, $T_e$([\ion{O}{iii}])$~=14000\pm650$\,K. We then assumed a two-zone approximation to define the temperature structure of the nebula, using $T_e$([\ion{O}{iii}]) as representative of high-ionization potential ions. The electron temperature assumed for the low-ionization potential ions was derived from the linear relation between $T_e$([\ion{O}{iii}])  and $T_e$([\ion{O}{ii}]) provided by \citet{G92}. We finally derived the ionic and total abundances of O, N, S, Ar and Ne, as well as the N/O, S/O, Ar/O and Ne/O ratios, for this \HII\ complex following the direct method. The results are compiled in Table~\ref{tab:abun}. In particular, we find a very high excitation degree in the gas, $\log$(O$^{++}$/O$^+$)\,$=0.78\pm0.09$, and derive \abox\,$=8.01\pm0.05$ and \lno\,$=-1.43\pm0.06$. These values are typical for \HII\ regions in BCDGs \citep[e.g.,][]{IT99,Izotov04,LSE10}.

We have combined all of the spaxels of the galaxy, excluding those belonging to the \HII\ complex, to quantify its global extinction and metallicity. However, now only the brightest emission lines are detected (see Table~\ref{tab:lines}), and hence we have to resort to strong-line methods (SEL) to determine the oxygen abundance of the gas. Table~\ref{tab:emp} compiles the results for the metallicity of the ionized gas following the most-common SEL techniques. For comparison, Table~\ref{tab:emp} also lists the oxygen abundance derived in the \HII\ complex following the same empirical methods. Besides the problem of the absolute abundance scale \citep[see][for details]{LSD12}, it is clear that the \HII\ complex has oxygen abundances which are systematically $\sim0.2$~dex lower than those observed in the rest of the galaxy. Assuming an oxygen abundance of \abox\,$=8.18$ (the average obtained using the \Te-based SEL methods), we computed electron temperatures, ionic and total abundances for the rest of the galaxy, which are compiled in Table~\ref{tab:abun}. Using the same method as used on the \HII\ complex, for the rest of the galaxy we calculate \chb\,$=0.31\pm0.04$ and \Wabs\,$=0.8\pm0.2$~\AA\ . 

$72$~per~cent of all the \Ha\ emission of GAMA\,J141103.98-003242.3 is found in the \HII\ complex. Using the extinction-corrected \Ha\ flux we derive a total \Ha\ luminosity of ($18.5\pm5.2$)$\times10^{39}$\,erg\,s$^{-1}$ for the \HII\ complex. Applying the  \cite{Ken98} relationship under the assumption of a \cite{Salpeter55} stellar initial mass function, we estimate a SFR of $0.147 \pm 0.041$\,M$_{\odot}$\,yr$^{-1}$. Using an O7V \Ha\ luminosity of $1.36\times10^{37}$\,erg\,s$^{-1}$\ from \citet{SV98}, the number of equivalent O7V stars needed to explain the \Ha\ luminosity is $1360$; similar to the number estimated for 30~Doradus \citep{Doran13}. Using prescriptions from \citet{Dia98}, the mass of ionized gas, $\log$(\MHII/\Mo)\,$=5.58\pm0.11$, and the mass of the stellar ionizing cluster, $\log$(\Mion/\Mo)\,$=5.81$, also matches those values found in 30~Doradus \citep{Faulkner67,Bosch09}. Assuming an instantaneous burst with $Z=0.008$ and considering the Starburst~99 models \citep{L99}, the \Ha\ equivalent width indicates that the last star-formation event in this \HII\ complex happened $4.8$\,Myr ago. These values are tabulated in Table~\ref{tab:Summ}. 

The combined \Ha-SFR of the galaxy including the \HII\ complex is $0.21\pm0.06$\,\Moy, which is higher than that measured by GAMA, though lower than the $FUV$-SFR measured from GALEX. It has been observed that the $FUV$-SFR is usually higher than the \Ha-SFR in dwarf galaxies \citep[e.g.][]{Lee09}. The discrepancy with the GAMA SFR could be explained by an inadequate aperture correction, as a $2$~arcsec fibre aperture placed on GAMA\,J141103.98-003242.3 would miss its \HII\ complex.

%%%%%%%%%%%%%%%%%%%%%%%%%%%%%%%%%%%%%%%%%%%%%%%%
\vspace{-0.4cm}
\section{Discussion \& Conclusions}

The analysis of GAMA\,J141103.98-003242.3 has revealed a luminous \HII\ complex with a SFR high enough to create a massive star cluster. This system has interesting implications for our understanding of dwarf galaxy star formation and evolution in the Local Universe. This galaxy is both visually and spectroscopically similar to the LMC-30~Doradus system, in hosting a single dominant star forming region in the outer disc, and in many ways could be considered as a more distant analogue of one of our nearest neighbours. 

Off-centered bright star-forming regions are often found in BCDGs \citep[e.g.][]{Loo86,Cai01,Gil03,LSE08}, but as we have shown in this paper, with IFU data we obtain detailed chemical and kinematic information. This allows us to probe the physical mechanisms that can explain the presence of this \HII\ complex in GAMA\,J141103.98-003242.3; either by external triggering or by the intrinsic stochasticity of star formation in an undisturbed system. With the data available there is no indication that the onset of star formation has been induced by an interaction with a companion. In particular there appear to be no kinematic disturbances in the ionised gas in the galaxy. The smooth kinematic distribution and low velocity dispersion at the location of the \HII\ complex are suggestive it is located in an undisturbed, rotationally supported disc. 

The main difference between GAMA\,J141103.98-003242.3 and the LMC is its isolation. GAMA assigns it one companion, GAMA\,J141120.29-002950.8, which is $\sim5$~arcmin, to the North-East and a physical separation of $\sim4$\,Mpc. There is no other catalogued companion (GAMA or SDSS) within 300\,km\,s$^{-1}$ and a 20~arcmin radius of GAMA\,J141103.98-003242.3. Visual inspection of optical (SDSS, $r\lesssim23$) and NIR (2MASS, K$_{s}\lesssim13.5$) imaging for this region revealed no indication of companion galaxies similar to or larger than GAMA\,J141103.98-003242.3.

The HIPASS data revealed a significant amount of \HI\ in the vicinity of GAMA\,J141103.98-003242.3, indicating \Mg/\Ms$~\sim11.2$, with $\log$(\MHI/\Mo)$~\sim9.5$ \citep[LMC $\log$(\MHI/\Mo)$~\sim8.7$;][]{Bru05}. Thus, the galaxy potentially has access to a large reservoir of neutral gas. The apparent lower metallicity of the \HII\ complex may suggest that the gas feeding the star formation has fallen in from the outer parts of the \HI\ disc following its expulsion during a previous episode of activity. Alternatively, the gas may have been supplied by an interaction with a companion galaxy as has been seen previously in other DG systems \citep[e.g.][]{LSK+12}, although the gas fraction with respect to the stellar mass and metallicity is not abnormal for similar galaxies \citep{Hua12,Hug13}. A lower mass companion that is less obvious could play a part, as seen in Tol 30 where an \HI\ tail extends towards a smaller dwarf galaxy \citep[L{\'o}pez-S{\'a}nchez et al. \emph{in prep}]{LS+10}. With no obvious companion near GAMA\,J141103.98-003242.3, and the fact that any galaxy too faint to be detected in the SDSS image would have an even more extreme gas/star ratio, the origin of the \HI\ gas from a previous interaction is both unlikely and unnecessary. Only with resolved \HI\ mapping of the region around GAMA\,J141103.98-003242.3 can a conclusion of its role on the \HII\ complex be made. 

The use of the term `\HII\ complex' in this work highlights the unresolved nature of this region, which warrants caution when deriving physical quantities. \citet{Ple00} showed that the \Ha\ flux of \HII\ regions can contribute up to $\sim40$~per~cent of the total \Ha\ flux of a nearly well-resolved galaxy, but up to $\sim75$~per~cent in the case where the spatial smoothing has been applied. This occurs because what used to be resolved \HII\ regions have now become an \HII\ complex due to the smoothing. Due to this, such unresolved \HII\ complexes are likely host to a collection of smaller \HII\ regions, but it has been found that it is more likely for the convolution of smaller \HII\ regions to produce an even light distribution instead of a point source distribution appearing as a single \HII\ region \citep{Elm09,Elm14}. If dwarf galaxies with intense star forming regions are remnants of the same evolutionary phase that clump-clusters went through, it is not unrealistic for this \HII\ complex to be dominated by a single \HII\ region (alike to 30~Doradus). With clump-clusters' growth being dominated not by merging activity, but rather smooth gaseous inflow \citep{Elm09}, the high gas fraction and isolation of GAMA\,J141103.98-003242.3 strengthens the scenario of stochastic gravitational collapse as opposed to interaction triggered star formation. 

The \HII\ complex has a measured \Ha\ luminosity of $10^{40.27}$\,erg\,s$^{-1}$, velocity dispersion $\simeq25$\,\kms and a diameter of $\lesssim600$\,pc. This agrees with the scaling relations found by \citet{Wis12} for high-$z$ $\sim$${\rm L}_{*}$ galaxies (e.g. clump-clusters; velocity dispersion $\propto$ radius $\propto$ \Ha\ luminosity $\propto$ Jeans mass), even though the \HII\ complex analysed here is hosted in a dwarf galaxy. In comparison to the work by \citet{Bau13}, the SFR and SSFR of GAMA\,J141103.98-003242.3 including the \HII\ complex are average for other galaxies of similar stellar mass. However, the SFR and SSFR of GAMA\,J141103.98-003242.3 excluding the \HII\ complex are at the extreme lower bounds for the same mass bin. This hints at a case where the star formation in such dwarf galaxies are typically dominated by a similar, single \HII\ complex. Taking into account GAMA\,J141103.98-003242.3's high gas fraction, we conclude that even though this \HII\ complex is significant for the galaxy itself (contributing $72$~per~cent of the total SFR of the galaxy), it is not currently experiencing a burst in its SFH where most of its stellar mass is formed. This is consistent with the work of \citet{Wei11}, which found that dwarf galaxies formed their underlying stellar population (accounting for $\sim85$~per~cent of their stellar mass) prior to $z=1$, with the rest of the stellar mass being formed by younger starbursts happening over the last 1 billion years.

These findings lead to the questions: 1. Where do the intense star forming regions of dwarf galaxies ($z\sim0.001$--$0.1$) lie on known scaling relations of clump-clusters? 2. What fraction of dwarf galaxies are undergoing extreme star formation in localised \HII\ complexes and what is their duty cycle? These questions are outside the scope of this paper, but suit well to future analysis of a sample of dwarf galaxies obtained with instruments such as SAMI\footnote{SAMI: http://sami-survey.org/}, KMOS\footnote{KMOS: http://www.eso.org/sci/facilities/develop/instruments/kmos.html} and MaNGA\footnote{MaNGA: http://www.sdss3.org/future/manga.php}. Considering the SAMI Galaxy Survey will observe $\sim400$ dwarf galaxies over the next few years, it will be possible to rigorously test this sample for more galaxies like GAMA\,J141103.98-003242.3, in the pursuit of understanding the star formation history of dwarf galaxies in the local universe and their place in the downsizing of star formation.

%%%%%%%%%%%%%%%%%%%%%%%%%%%%%%%%%%%%%%%%%%%%%%%%
\vspace{-0.4cm}
\section{Acknowledgments}

The SAMI Galaxy Survey is based on observation made at the Anglo-Australian Telescope. The Sydney-AAO Multi-object Integral field spectrograph (SAMI) was developed jointly by the University of Sydney and the Australian Astronomical Observatory. The SAMI input catalogue is based on data taken from the Sloan Digital Sky Survey, the GAMA Survey and the VST ATLAS Survey. The SAMI Galaxy Survey is funded by the Australian Research Council Centre of Excellence for All-sky Astrophysics (CAASTRO), through project number CE110001020, and other participating institutions. The SAMI Galaxy Survey website is http://sami-survey.org/.

The ARC Centre of Excellence for All-sky Astrophysics (CAASTRO) is a collaboration between The University of Sydney, The Australian National University, The University of Melbourne, Swinburne University of Technology, The University of Queensland, The University of Western Australia and Curtin University, the latter two participating together as the International Centre for Radio Astronomy Research (ICRAR). CAASTRO is funded under the Australian Research Council (ARC) Centre of Excellence program, with additional funding from the seven participating universities and from the NSW State Government's Science Leveraging Fund.

Funding for SDSS-III has been provided by the Alfred P. Sloan Foundation, the Participating Institutions, the National Science Foundation, and the U.S. Department of Energy Office of Science. The SDSS-III web site is http://www.sdss3.org/.

GAMA is a joint European-Australasian project based around a spectroscopic campaign using the Anglo-Australian Telescope. The GAMA website is http://www.gama-survey.org/.

The Parkes telescope is part of the Australia Telescope which is funded by the Commonwealth of Australia for operation as a National Facility managed by CSIRO.

SMC acknowledges support from an ARC Future fellowship (FT100100457). ISK is the recipient of a John Stocker Postdoctoral Fellowship from the Science and Industry Endowment Fund (Australia). JTA acknowledges the award of an ARC Super Science Fellowship (FS110200013). MSO acknowledges the funding support from the Australian Research Council through a Super Science Fellowship (ARC FS110200023).

We would also like to thank the anonymous reviewer for their time and effort in the thorough checking of this paper.

%%%%%%%%%%%%%%%%%%%%%%%%%%%%%%%%%%%%%%%%%%%%%%%%

%%%%%%%%%%%%%%%%%%%%%%%%%%%%%%%%%%%%%%%%%%%%%%%%

\section{Appendix}

Table~\ref{tab:lines} compiles the dereddened line intensity ratios with respect to $I$(\,\Ha\,)\,$=100$. Table~\ref{tab:abun} lists the results of the chemical abundance analysis for the \HII\ complex discovered in the galaxy GAMA\,J141103.98-003242.3 and the center of the galaxy. Table~\ref{tab:emp} compiles the oxygen abundances derived using the most commonly used strong emission-line methods.

%%%%%%%%%%%%%%%%%%%%%%%%%%%%%%%%%%%%%%%%%%%%%%%%
\vspace{5cm}
\begin{table}
\caption{Dereddened line intensity ratios with respect to $I$(H$\beta$)=100 for the \HII\ complex discovered in the galaxy GAMA\,J141103.98-003242.3. The ionized gas observed in the rest of the galaxy is also presented. At the bottom of the table we also give the \Hb\ flux, the  reddening coefficient, $c$(\,\Hb\,), the equivalent widths of the absorption in the hydrogen lines, $W_{abs}$, and the equivalent widths of the emission \ion{H}{i} Balmer lines. The value of $f(\lambda)$ considering the  \citet*{Cardelli89} extinction law and used for dereddening the line intensity ratios are also included. A colon denotes an error of larger than 40\%. \label{tab:lines}}
\scriptsize
{\centering

\begin{tabular} {l r c c}
\hline

 Line                      & $f(\lambda)$ &   \HII\ complex  & Rest of the galaxy  \\

\hline

% [\ion{S}{iii}]~3722	        & $0.324$   &     $ 7.21\pm0.68$   &  \nodata    \\
 
 [\ion{O}{ii}]~3728& $0.322$   &      $91.3\pm5.6 $   &  $352\pm52$  \\ 
 
 H11~3770                       & $0.313$   &      $5.52\pm0.42$	 &   \nodata     \\  
 
 [\ion{Ne}{iii}]~3869 		& $0.291 $  &      $60.4\pm3.8	$ &   \nodata    \\  
 
 HeI 3889 + H8                        & $0.286  $ &      $20.9\pm2.2$	 &   \nodata    \\  
 
 [\ion{Ne}{iii}]~3969~+~H7	& $0.267 $  &     $ 36.7\pm3.6$	 &   \nodata   \\  
 
 H$\delta$~4101 		& $0.230 $  &      $26.1\pm2.0	$ & $25.0:$  \\  
 
 H$\gamma$~4340 		& $0.157 $  &      $47.5\pm2.3$	 &   $43\pm6.7$     \\  
 
 [\ion{O}{iii}]~4363 	 	& $0.150$   &      $11.4\pm0.8$	 &   \nodata     \\  
 
 \ion{He}{i}~4471   		& $0.116 $  &      $3.87\pm0.45$   &   \nodata      \\  
 
 \ion{He}{ii}~4686     & $0.050 $  &      $0.51:$   &   \nodata      \\  
 
 [\ion{Ar}{iv}]~+~\ion{He}{i}~4712 		& $0.043 $  &      $1.72\pm0.32$	 &   \nodata     \\
 
 [\ion{Ar}{iv}]~4740 		& $0.034 $  &      $0.99:$		 &  \nodata     \\
 
 H$\beta$~4861    		& $0.000$   &    $100.0\pm3.6$	 &  $100\pm14$  \\
 
 [\ion{O}{iii}]~4959 		&$-0.025$ &    $234\pm11$	 &  $168\pm23$ \\
 
 [\ion{O}{iii}]~5007 		&$-0.037$ &    $695\pm30$	 &  $482\pm58$  \\
 
 [\ion{O}{i}]~6300  		&$-0.262$ &    $1.70\pm0.46$ &   \nodata    \\
 
 [\ion{S}{iii}]~6312  		&$-0.264$ &      $1.73\pm0.79$  &   \nodata     \\
 
  [\ion{O}{i}]~6364             &$-0.271$ &      $0.78\pm0.11$		 &   \nodata     \\

[\ion{N}{ii}]~6548   		&$-0.295$ &      $1.75\pm0.17$	 &   $6.25:$   \\

H$\alpha$ 6563   		&$-0.297$ &       $279\pm12$	 &  $281\pm34$ \\

[\ion{N}{ii}]~6583 		&$-0.300$ &      $4.88\pm0.29$	 &  $16.7\pm2.1$ \\

 \ion{He}{i}~6678   		&$-0.312$ &      $3.17\pm0.35$	 &   \nodata   \\ 

 [\ion{S}{ii}]~6716  		&$-0.318$ &      $7.67\pm0.44$	 &  $38.9\pm4.7$ \\

 [\ion{S}{ii}]~6731   		&$-0.319$ &      $6.01\pm0.35$	 &  $24.6\pm3.9$ \\

 \ion{He}{i}~7065  		&$-0.364$ &      $2.74\pm0.42$	 &   \nodata    \\

 [\ion{Ar}{iii}]~7135		&$-0.373$ &      $10.33\pm0.57$	 &  $20.9\pm3.8$  \\

 \hline

 \noalign{\smallskip}                   
                    
EW(H$\alpha$) [\AA] & &       $451\pm11$	& $34.8\pm3.2$ \\
EW(H$\beta$)  [\AA] & &       $137\pm4$	& $6.8\pm1.6$\\
EW(H$\gamma$) [\AA] & &      $44.2\pm1.6$	&   $1.7\pm0.4$      \\
EW(H$\delta$) [\AA] & &      $22.8\pm0.9$    &   $1.1:$    \\

 \hline

\noalign{\smallskip}

${\emph{F}}_{\rm H\beta}$$^{\Diamond}$  & &     $22.6\pm6.3^{\dagger}$   &  $9.4\pm2.6^{\dagger}$    \\

$c$(\,\Hb\,)         & &    $0.34\pm0.02$    &     $0.31\pm0.04$   \\
$W_{abs}$   [\AA]     & &         $0.0\pm0.1$   &    $0.8\pm0.2$  \\

\hline
   
\end{tabular}\\}

$^{\Diamond}$ Units of 10$^{-16}$\,erg\,cm$^{-2}$\,s$^{-1}$\\
$^{\dagger}$ Error is currently dominated by a $28$~per~cent error in SAMI's absolute flux calibration, and in all values derived from it (see \citet{All14} for more details). \\

\end{table}

%%%%%%%%%%%%%%%%%%%%%%%%%%%%%%%%%%%%%%%%%%%%%%%%

\begin{table}
\caption{Physical conditions and chemical abundances of the ionized gas for the \HII\ complex discovered in the galaxy GAMA\,J141103.98-003242.3 and the rest of the galaxy when this region is not considered. In this latter case, the electron temperatures were estimated as those that best reproduce the oxygen abundance computed via the SEL methods based on \Te. Bold values are those of most interest. \label{tab:abun}}
\scriptsize
\centering

\begin{tabular} {l c c}
\hline

          &   \HII\ complex  & Rest of the galaxy \\

\hline

$T_e$[\ion{O}{iii}]  [K]  &  $14000\pm650$    &     $12750\pm1000$ \\
$T_e$[\ion{O}{ii}]   [K]  &  $12800\pm450$    &     $11930\pm800$ \\
$n_e$ [cm$^{-3}$]         &   $140\pm30$       &       $100$ \\

\noalign{\smallskip}

12+log(O$^+$/H$^+$)       &  $7.16\pm0.07$ &   $7.85\pm0.13$ \\
12+log(O$^{++}$/H$^+$)    &  $7.94\pm0.05$ &   $7.90\pm0.09$ \\    
{\bf 12+log(O/H)   }      &  $\bf 8.01\pm0.05$ &  $\bf 8.18\pm0.11$ \\ 
log(O$^{++}$/O$^+$)       &  $0.78\pm0.09$ &   $0.05\pm0.15$ \\	     
										  
\noalign{\smallskip}								  
										  
12+log(N$^+$/H$^+$)       &  $5.74\pm0.05$  &      $6.35\pm0.08$  \\		 
12+log(N/H)               &  $6.58\pm0.09$  &      $6.67\pm0.11$   \\	 
log(N/O)                  & $-1.43\pm0.06$  &     $-1.51\pm0.10$  \\ 		 
										  
\noalign{\smallskip}								  
										  
12+log(S$^+$/H$^+$)      &  $5.27\pm0.04$	& $5.99\pm0.07$ \\ 			  
12+log(S$^{++}$/H$^+$)   &  $6.06\pm0.18$   &  \nodata \\				     
12+log(S/H)              &  $6.27\pm0.16$	&  \nodata  \\  		   
log(S/O)                 & $-1.74\pm0.21$   &  \nodata \\				 
										  
\noalign{\smallskip}								  
										  
12+log(Ne$^{++}$/H$^+$)  &   $7.29\pm0.06$ &  \nodata \\		  
12+log(Ne/H)             &   $7.36\pm0.12 $ &  \nodata \\  	    
log(Ne/O)                &  $-0.65\pm0.07$ &  \nodata \\		    
				   					  
\noalign{\smallskip}								  
12+log(Ar$^{++}$/H$^+$)  &  $5.67\pm0.06$ &	 $6.06\pm0.12$ \\				    
12+log(Ar$^{+3}$/H$^+$)  &  $5.00\pm0.11$  &  \nodata \\				    
12+log(Ar/H)             &  $5.77\pm0.07$	 & $5.90\pm0.20$ \\
log(Ar/O)                & $-2.24\pm0.12$   & $-2.29\pm0.22$ \\

 \hline

\noalign{\smallskip}

$c$(\,\Hb\,)         &    $0.34\pm0.02$    &     $0.31\pm0.04$  \\
$W_{abs}$   [\AA]     &         $0.0\pm0.1$   &    $0.8\pm0.2$  \\

\hline
   
\end{tabular}

\end{table}

%%%%%%%%%%%%%%%%%%%%%%%%%%%%%%%%%%%%%%%%%%%%%%%%

\begin{table*}
\caption{Oxygen abundances derived using the most commonly used strong emission-line methods. The strong emission-line calibrations are: 
 M91: \citet{McGaugh91};
 KD02: \citet{KD02}; 
 KK04: \citet{KK04}
 PT05: \citet{PT05};  
 P01: \citet{P01a,P01b}; 
 PP04a: \citet{PP04}, 
 using a linear fit to the $N_2$ parameter;  
 PP04c:  \citet{PP04}, using the $O_3N_2$ parameter.
 The last two columns list the average abundance value using all the empirical methods, the $T_e$ method is not considered here. We provide two results: PPP, which considers the average value obtained with the  PT05, P01, PP04a  and  PP04c calibrations and MKD, which assumes the average value of the M91, KD02, and KK04 calibrations. The typical  uncertainty in these values is $\sim0.10$\,dex. Bold values are those of most interest.
 \label{tab:emp} }
\scriptsize

 \begin{tabular}{l  cc c ccc  cccc  cc c}
\hline

   & $c$(\,\Hb\,) & $W_{abs}$ [\AA] & $T_e$     &  M91   & KD02  & KK04     &     PT05  &  P01   &     PP04a   &    PP04c        &  \multicolumn{2}{c@{\hspace{4pt}}}{Adopted}   & Branch \\

\noalign{\smallskip}
\cline{10-11}
\noalign{\smallskip}

Parameters            & &       &   &    $R_{23}$, $y$  & $R_{23}$, $y$   &    $R_{23}$, $y$&         $R_{23}$, $P$  & $R_{23}$, $P$ &
                          $N_2$    &  $N_2O_3$   &             MKD     &    PPP   \\

\hline 

\HII\ complex  & $0.34\pm0.02$ & $0.0\pm0.1$ & {$\bf 8.01\pm0.05$} &    $8.17$  &  $8.33$ & $8.35$ &  $7.99$ & $7.88$ &  $7.90$ &    $7.90$ &    {$\bf 8.28$}  & {$\bf 7.92$} & Low   \\
Rest of the galaxy    & $0.31\pm0.04$ &   $0.8\pm0.2$  & \nodata    &    $8.35$  &  $8.51$ & $8.52$ &   $8.25$ & $8.13$ & $ 8.20$ &  $  8.12$ &    {$\bf 8.45$}  & {$\bf 8.18$} & Intermediate  \\
{\it Difference}    &   \nodata      &   \nodata &   \nodata            &    $0.18$  &  $0.18$  & $0.17$ &   $0.26$ & $0.25$ & $0.30$ &   $ 0.22 $&     {$\bf 0.18$}  &  {$\bf 0.26$} & \nodata \\

\hline 
\end{tabular}

\end{table*}

%%%%%%%%%%%%%%%%%%%%%%%%%%%%%%%%%%%%%%%%%%

\end{document}